\documentclass[12pt,preprint]{aastex}

\newcommand\rhodm{\rho_{\rm DM}}
\newcommand\calL{\mathcal{L}}
\newcommand\ksp{\kappa_{\rm sp}}
\newcommand\calH{\mathcal{H}}
\newcommand\Fc{F_{\rm c}}

\begin{document}

\title{Equilibrium Models of Galaxy Clusters with Cooling, Heating and Conduction}

\author{M.Br\"{u}ggen\altaffilmark{1,2}}
\altaffiltext{1}{International University Bremen, Campus Ring 1, 28759 Bremen, Germany}
\altaffiltext{2}{Max-Planck Institut f\"ur Astrophysik, Karl-Schwarzschild-Str 1, 85740 Garching, Germany}
\email{m.brueggen@iu-bremen.de}

\begin{abstract}
It is generally argued that most clusters of galaxies host cooling
flows in which radiative cooling in the centre causes a slow
inflow. However, recent observations by Chandra and XMM conflict with
the predicted cooling flow rates.  Amongst other mechanisms, heating
by a central active galactic nucleus and thermal conduction have been
invoked in order to account for the small mass deposition rates. Here,
we present a family of hydrostatic models for the intra-cluster medium
where radiative losses are exactly balanced by thermal conduction and
heating by a central source. We describe the features of this simple
model and fit its parameters to the density and temperature profiles
of Hydra A.
\end{abstract}




\section{Introduction}

The X-ray surface brightness of many clusters of galaxies shows a
strong central peak which has been interpreted as the signature of a
cooling flow (Cowie \& Binney 1977, Fabian \& Nulsen 1977, Sarazin
1988, Fabian 1994). However, the simple cooling flow model conflicts
with a growing number of observations that show that while the
temperature is declining in the central region, gas with a temperature
below $\sim$1 keV is significantly less abundant than predicted.  The
absence of a cool phase in cooling flows has proven a persistent
puzzle. Recently, two main candidates for the heating of the gas in
the central regions of clusters have emerged: (i) heating by outflows
from active galactic nuclei (AGN) (\citealt{tabo93}, Churazov et
al. 2001a, \citealt{ bin01}, \citealt{brueg02}, Br\"uggen 2003),
and (ii) transport of heat from the outer regions of the cluster by
thermal conduction (\citealt{zak03}, \citealt{rus02},
\citealt{fabi01}, \citealt{voig02}, \citealt{gruz02}, \citealt{fri86},
\citealt{bert86}, \citealt{meik88}, \citealt{breg88}).

The role of thermal conduction in the ICM has been the subject of a
long debate and, owing to the complex physics of MHD turbulence, the
value of the effective conductivity remains uncertain. The thermal
conductivity of an unmagnetised, fully ionised plasma was calculated
by \citet{spi62}. Originally it has been thought that the magnetic
field in clusters strongly supresses the thermal conductivity because
the magnetic fields prevent an efficient transport perpendicular to
the field lines. Even if the transport can be efficient along the
magnetic field lines, the overall isotropic conductivity was thought
to be many orders of magnitude less than the Spitzer value. This
paradigm has been supported by a number of observations, such
as sharp edges at so-called cold fronts, small-scale temperature
variations in mergers and sharp boundaries around radio bubbles. It is
thought that the existence of these sharp features precludes thermal
conduction near the Spitzer value (Markevitch et al. 2000, Vikhlinin,
Markevitch \& Murray 2001).

Recent theoretical work by \cite{nara01}, Malyshkin \& Kulsrud (2001),
Chandran et al. (1999), Chandran \& Cowley (1998) and earlier work by
Rechester \& Rosenbluth (1978) has shown that a turbulent magnetic
field is not as efficient in suppressing thermal conduction as
previously thought. It is argued that chaotic transverse motions of
the tangled magnetic field lines can enhance the cross-field diffusion
to an extent that the effective conductivity is of the order of the
Spitzer value. Following this work, \cite{zak03} have looked for
hydrostatically stable models in which the radiative cooling is
exactly balanced by thermal conduction and where the temperatures on
the inner and outer boundaries were fixed.  For half of the clusters
they investigated they found that a thermal conductivity of around 30
\% of the Spitzer value yielded good fits to the observed
profiles. However, for the other half of the clusters conduction alone
appeared unlikely to halt a cooling catastrophe. Consequently, some
form of heating is necessary such as mechanical heating by AGN. The
work by \cite{zak03} has motivated us to reexamine cluster models that
takes into account thermal conduction as well as heating by central
radio sources. We include a more generally valid treatment of
radiative cooling by utilising a fit to the cooling function that
takes into account line cooling.  This becomes particularly important
in the central regions of cooler clusters.\\

Radio-loud AGN drive strong outflows in the form of jets that inflate
bubbles or lobes. The lobes are filled with hot plasma, and can heat
the cluster gas in various ways. The effect of hot bubbles on the ICM
consists primarily of heating via $pdV$ work and redistribution of
mass via buoyancy-driven mixing. Hydrodynamic simulations (Br\"uggen
2003) have shown that a significant fraction ($\sim 10$ \%) of the
energy residing in radio lobes can be dissipated in the cluster gas.

Clearly, the lifetime of the activity of the central AGN (as brief as
$\sim 10^{4}-10^{5}$ yrs) is short compared to the evolutionary time
scale of the cluster gas. Therefore, once the AGN stops supplying
energy to the buoyant bubbles, the cluster gas will settle down once
again and a full cooling flow may be re-established. The cooling gas
flows to the centre of the cluster and may then trigger a further
active phase of the AGN. Thus a self-regulating process with cooling
periods alternating with brief bursts of AGN activity may be
established (\citealt{qui01}, \citealt{voit01}). The rising bubbles
uplift colder material from the vicinity of the AGN and thus disrupt
the supply of fresh fuel for the radio jets. This feedback mechanism
might automatically regulate the power of the radio jets. It has been
found that 71 \% of all cD galaxies at the centres of cooling-flow
clusters show evidence of radio activity (\citealt{bur90}). This
fraction is higher than in non-cooling flow galaxies which again
points towards the existence of some form of feedback mechanism where
the cooling gas flows to the centre of the cluster and triggers a
further active phase of the AGN. The bubbles rise at a speed that is
comparable to the sound speed, which in turn is comparable to the
dynamical speed. Therefore, the cooling timescale is much longer than
the bubble rise timescale, and it is justifiable to treat feedback
heating in a time-averaged sense.

In this paper, we will present a simple model that incorporates, both,
thermal conduction and heating by bubbles. We compute a class of
models that are subject to radiative cooling, thermal conduction,
heating by AGN and that are in hydrostatic equilibrium. We will
investigate those parameters that yield plausible solutions and will
compare our solutions to observations. Finally, we briefly discuss the
origin of the energy that is conducted into the cluster.

\section{Model}

In our model of the cluster, we assume that the dark matter
distribution is given by a modified NFW profile
(\citealt{nava97})

\begin{equation}
\rhodm (r)=\frac{M_0/2\pi}{(r+r_c)(r+r_s)^2},
\end{equation} 
where $r$ is the distance from the centre, $r_s$ the scale radius of
the NFW profile and $r_c$ a softening radius, within which the density
becomes constant and which prevents the temperature from going to zero
at $r=0$. The core radius $r_c$ determines the shape of the potential
near the centre and here we adopt $r_c=r_s/20$ as recommended by
\citet{zak03}. We can express the characteristic mass $M_0$ in terms
of the commonly used concentration parameter $c=(3M_{\rm vir}/4\pi
200\rho_{\rm crit}(z)r_c^3)^{1/3}$, where $\rho_{\rm crit}$ is the
critical density of the universe at the redshift of the cluster and
$M_{\rm vir}$ is the virial mass:

\begin{equation}
M_0=2\pi r_s^2 r_c\rho_{\rm crit}(z)\left (\frac{200}{3}\right
)\frac{c^3}{\ln (1+c)-c/(1+c)} \ .
\end{equation}
The conductive heat flux is given by

\begin{equation}
\Fc=-\nabla\cdot (\kappa\nabla T),
\end{equation}
where $\kappa$ is the thermal conductivity and $T$ temperature. If
thermal conduction is due to electrons, the conductivity according to
Spitzer (1962) is given by

\begin{equation}
\ksp \approx 9.2\ 10^{-7} T^{5/2} {\rm erg}\ {\rm s}^{-1}\
{\rm K}^{-1}\ {\rm cm}^{-1}.
\end{equation}
Here we seek an equilibrium model that is spherically symmetric,
static, and time-independent. Moreover, we neglect magnetic fields and
the self-gravity of the ICM. In spherical coordinates, the
gravitational potential $\Phi$ is governed by the dark matter
distribution

\begin{equation}\label{Phi}
\frac{1}{r^2}\frac{d}{dr}\left(r^2\Phi\right)
= 4\pi G\rhodm(r) 
=\frac{2GM_0}{(r+r_c)(r+r_s)^2} \ ,
\end{equation}
where $G$ is the gravitational constant. Following \cite{zak03} the
mass-temperature relation of \citet{afs02} and the mass-scale relation
of \citet{mao97}, can be used to determine $M_0$ and $r_s$ for a given
cluster. The equation of hydrostatic equilibrium can now be written as

\begin{equation}\label{P}
\frac{dp}{dr} = - \rho \frac{d\Phi}{dr} \ ,
\end{equation}
where $p$ is pressure and $\rho$ gas density. Demanding that radiative
cooling and heating are balanced by thermal conduction, we can write

\begin{equation}\label{Fr}
\frac{1}{r^2}\frac{d}{dr}\left(r^2\Fc\right) = -\rho\calL +\calH \ ,
\end{equation}
where $\rho\calL$ and $\calH$ are the volume cooling and heating
functions, respectively, and $\Fc$ is the conductive flux, which
depends on the temperature gradient as stated above

\begin{equation}\label{T}
\kappa\frac{dT}{dr} = -\Fc \ ,
\end{equation}
where we write $\kappa$ as $f\ksp$, $f$ being a suppression factor.  We
should point out that we omitted a term for the convective flux in
equation (\ref{Fr}). However, for the levels of heating considered here
the ICM is convectively stable and the neglect of convection is a
justifiable assumption.

Pressure is related to $n_e$ and $T$ via the ideal gas law

\begin{equation}\label{eos}
p = \frac{\rho k T}{\mu m_u} = \frac{\mu_e}{\mu} n_e k T,
\end{equation}
where $m_u$ is the atomic mass unit.\\

For the volume cooling rate $\rho\calL$ we use an approximation to the
cooling function based on calculations by \citep{sd93}

\begin{equation}
\rho\calL =[C_{1}(k T)^{\alpha}+C_{2}(k T)^{\beta}+C_{3}]\frac{\mu_e}{\mu} n_e^2\ 10^{-22}\; {\rm erg}\; {\rm cm}^{3}\; {\rm s}^{-1} \ .
\end{equation}
The units for $kT$ are keV, $n_e$ is the electron number density,
and $\mu$ and $\mu_e$ denote the mean molecular weight per hydrogen
atom and per electron, respectively. As in \cite{zak03} we use
$\mu=0.62$ and $\mu_e=1.18$, corresponding to a fully ionized gas with
hydrogen fraction $X=0.7$ and helium fraction $Y=0.28$. For an average
metallicity $Z=0.3 Z_{\odot}$ the constants are $\alpha =-1.7$, $\beta
=0.5$, $C_1=8.6\times 10^{-3}$, $C_2=5.8\times 10^{-2}$ and
$C_3=6.4\times 10^{-2}$.\\

The term $\calH$ in equation (\ref{Fr}) represents the energy input by a
central AGN. The energy is deposited in the ICM by the rising bubbles
and thus, averaged over time, the heating will be distributed in
radius. To quantify this heating, we use a prescription proposed by
\cite{be01} which quantifies the heating of the ICM by the
rising bubbles. Assuming that this heating mechanism reaches a
quasi-steady state, the details of the bubble motions and geometry
should cancel and the energy flux may be written as

\begin{equation}
\dot{e}\propto p_{\rm b}(r)^{(\gamma_{\rm b}-1)/\gamma_{\rm b}},
\end{equation}
where $p_{\rm b}(r)$ is the partial pressure of buoyant gas inside
bubbles at radius $r$ and $\gamma_{\rm b}$ is the adiabatic index of
buoyant gas which we will take to be 4/3.  Assuming that the partial
pressure scales with the thermal pressure of the ICM, the volume
heating function $\calH$ can be expressed as

\begin{equation}
\calH \sim -h(r)\nabla\cdot\frac{\dot{e}}{4\pi
r^{2}}=-h(r)\left(\frac{p}{p_0}\right)^{(\gamma_{b} -1)/\gamma_{b}}\frac{1}{r}\frac{d\ln p}{d\ln r},
\end{equation}
where $p_0$ is the central pressure and $h(r)$ is the normalization function

\begin{equation}\label{smallh}
h(r)=\frac{L}{4\pi r^{2}}(1-e^{-r/r_0})q^{-1},
\end{equation}
where $L$ is the luminosity of the AGN. The normalisation factor $q$
is defined by

\begin{equation}
q=\int_{r_{\rm min}}^{r_{\rm max}}\left(\frac{p}{p_0}\right)^{(\gamma_{b} -1)/\gamma_{b}}\frac{1}{r}\frac{d\ln p}{d\ln r}(1-e^{-r/r_0}) {\rm d}r \ ,
\end{equation}
where $r_0$ is the inner heating cutoff radius which is determined by
the finite size of the central radio source. Here $r_0$ was taken to
be 10 kpc. This heating function mimics a possible feedback mechanism
between the AGN and the ICM in the sense that the volume heating
function does not depend on the physical conditions at the source
alone but on the pressure gradient (\citealt{rus02}). Thus, it resembles
thermal conduction, with the difference that the heating rate depends
on the gradient of pressure rather than temperature.\\

Equations (\ref{Phi}) - (\ref{eos}) can be combined to yield a set of
four first-order differential equations for $n_e$, $r^2 d\Phi/dr$, $T$
and $r^2 \Fc$. We solve these equations as an initial-value problem
subject to the initial conditions: $n_e(0)=n_{\rm in}$, $r^2
d\Phi/dr|_{r=0} = 0$, $T(0)=T_{\rm in}$ and $r^2 \Fc|_{r=0}=0$, where
$n_{\rm in}$ and $T_{\rm in}$ are parameters. Other parameters in this
simplified model are $M_0$, $r_s$, $f$ and $L$. We have investigated
the dependence of the resulting density and temperature profiles on
some of these parameters, which is discussed in the next session.

\section{Results and Discussion}

\begin{figure}[htp]
\plotone{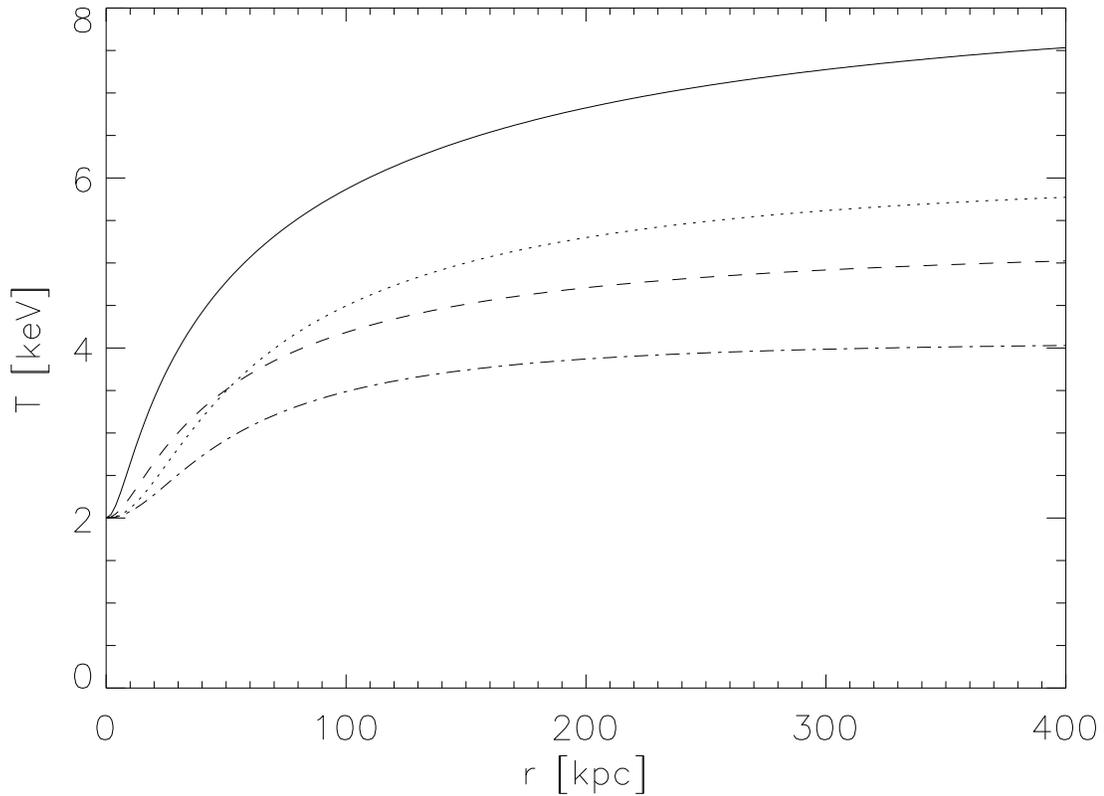} 
\caption{Temperature profiles for a cluster with central temperature
$T_{\rm in}=2$ keV and central density of $n_e=0.05$ cm$^{-3}$ with
$f=0.1$, $L=3\times 10^{43}$ erg s$^{-1}$ (solid) and $f=0.1$,
$L=3\times 10^{44}$ erg s$^{-1}$ (dotted), $f=0.5$, $L=3\times
10^{43}$ erg s$^{-1}$ (dashed) and $f=0.5$, $L=3\times 10^{44}$ erg s$^{-1}$
(dash-dot).}
\label{fig1}
\end{figure}

\begin{figure}[htp]
\plotone{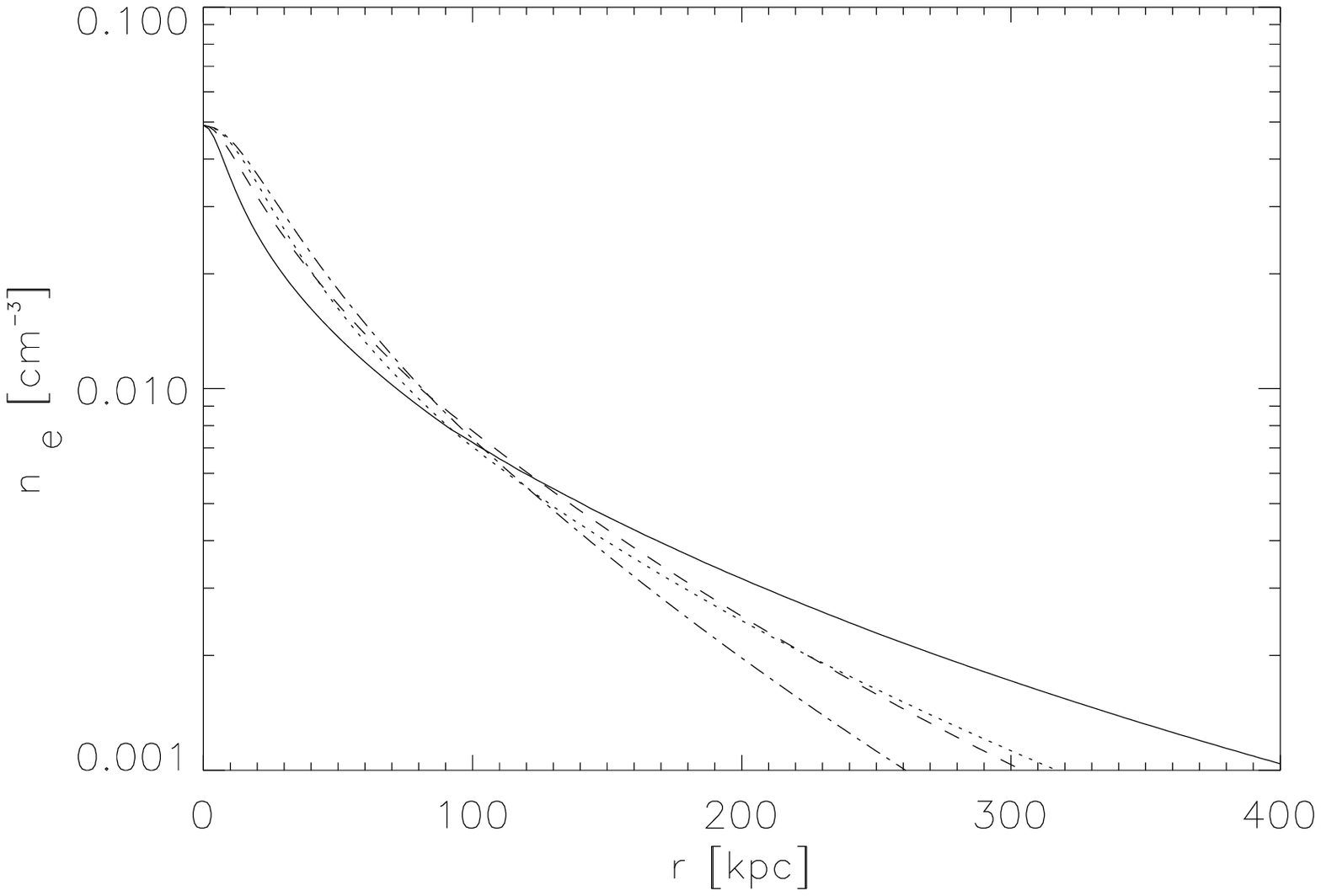} 
\caption{Electron number density profiles corresponding to the models presented in Fig. 1.}
\label{fig2}
\end{figure}

We integrate the system of four ordinary differential equations using
a Runge-Kutta method and adopt a characteristic mass of $M_0 = 6.6
\times 10^{14}\ M_{\odot}$ and a scale radius of $r_s=460$ kpc
(parameters inferred for the cluster Abell 1795, \citealt{zak03}). For
a cluster with a central temperature of 2 keV and a central density of
$n_e=0.05$ cm$^{-3}$ the resulting density and temperature profiles
are shown in Figures \ref{fig1} and \ref{fig2}. Curves plotted in
different styles correspond to different values of the suppression
factor $f$ and the luminosity of the central source $L$. The values
assumed for $L$ here correspond to typical energies supplied by the
jet to the ICM of $\sim 10^{44}$ erg s$^{-1}$ (e.g. Owen, Eilek and
Kassim 2000). In Figure \ref{fig1} one can note that the temperature
profile rises less steeply with radius if thermal conduction is more
efficient. In order to maintain a given central temperature, the
temperatures gradient in the cluster has to be higher if conduction is
more suppressed. The higher temperature gradient makes up for the
smaller suppression factor so that enough energy is conducted inwards
to balance the radiative losses. The highest value for the suppression
factor that we have adopted here (0.5) is at the upper end of what is
physically plausible. Even higher thermal conductivities seem very
unlikely. Moreover, one can see, that the heating term has an effect
similar to that of thermal conduction, in that a higher luminosity of
the central source flattens out the temperature profile. Because the
distributed heating depends on the pressure gradient, a higher
luminosity of the central source can afford a smaller pressure
gradient, leading to a shallower temperature distribution. Physically
speaking, the smaller pressure gradient makes the energy transfer from
the bubbles to the ICM less efficient.

For higher values of $f$ and $L$ the electron number density is higher
in the centre ($<$ 100 kpc) and drops off faster in the outer regions
(see Fig. \ref{fig2}).

\begin{figure}[htp]
\plotone{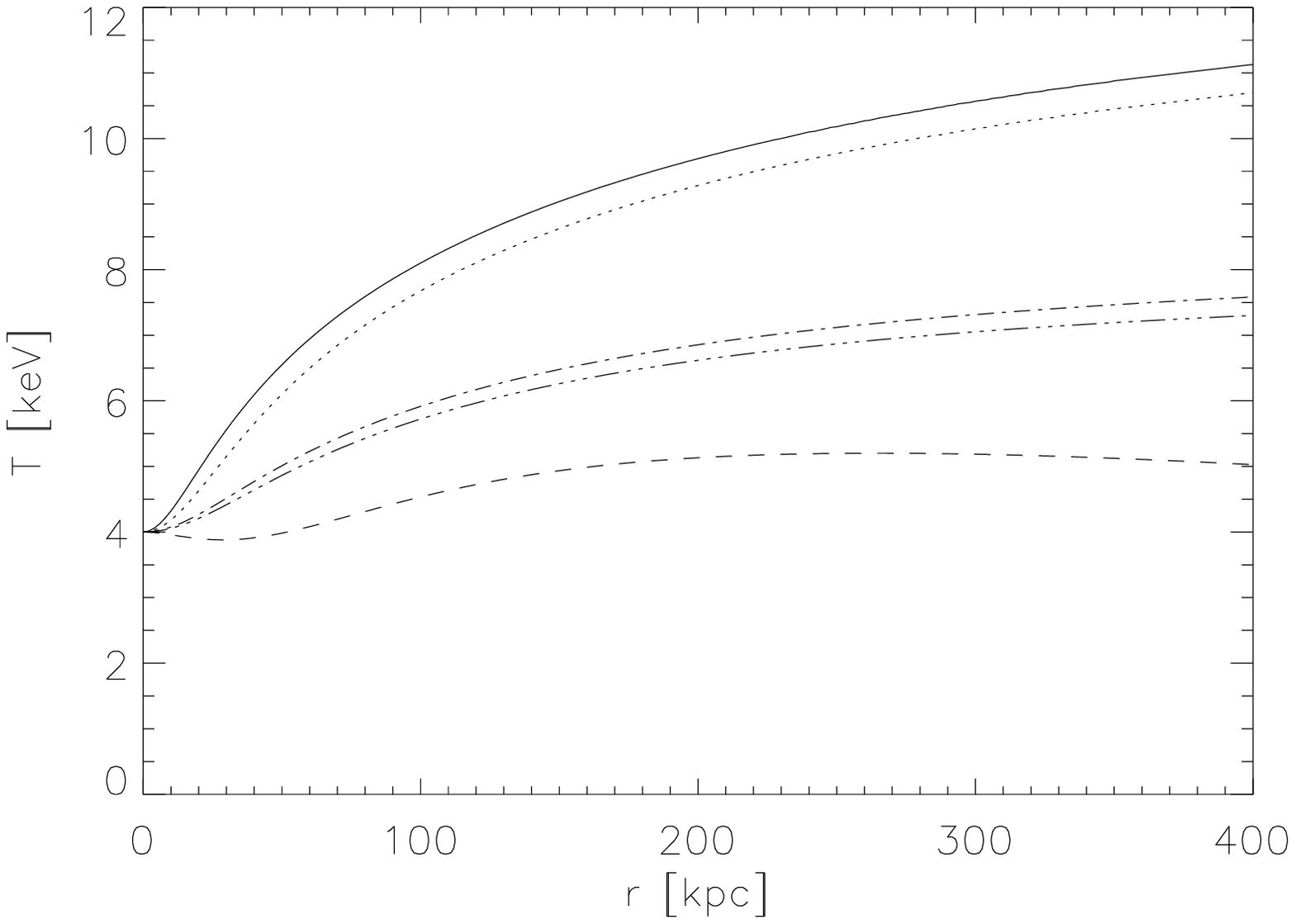} 
\caption{Temperature profiles for a cluster with central temperature
$T_{\rm in}=4$ keV and central density of $n_e=0.05$ cm$^{-3}$ with
$f=0.1$, $L=3\times 10^{43}$ erg s$^{-1}$ (solid) and $f=0.1$,
$L=3\times 10^{44}$ erg s$^{-1}$ (dotted), $f=0.1$, $L=3\times
10^{45}$ erg s$^{-1}$ (dashed), $f=0.5$, $L=3\times 10^{43}$ erg
s$^{-1}$ (dash-dot) and $f=0.5$, $L=3\times 10^{44}$ erg s$^{-1}$
(dash-dot-dot).}
\label{fig3}
\end{figure}

\begin{figure}[htp]
\plotone{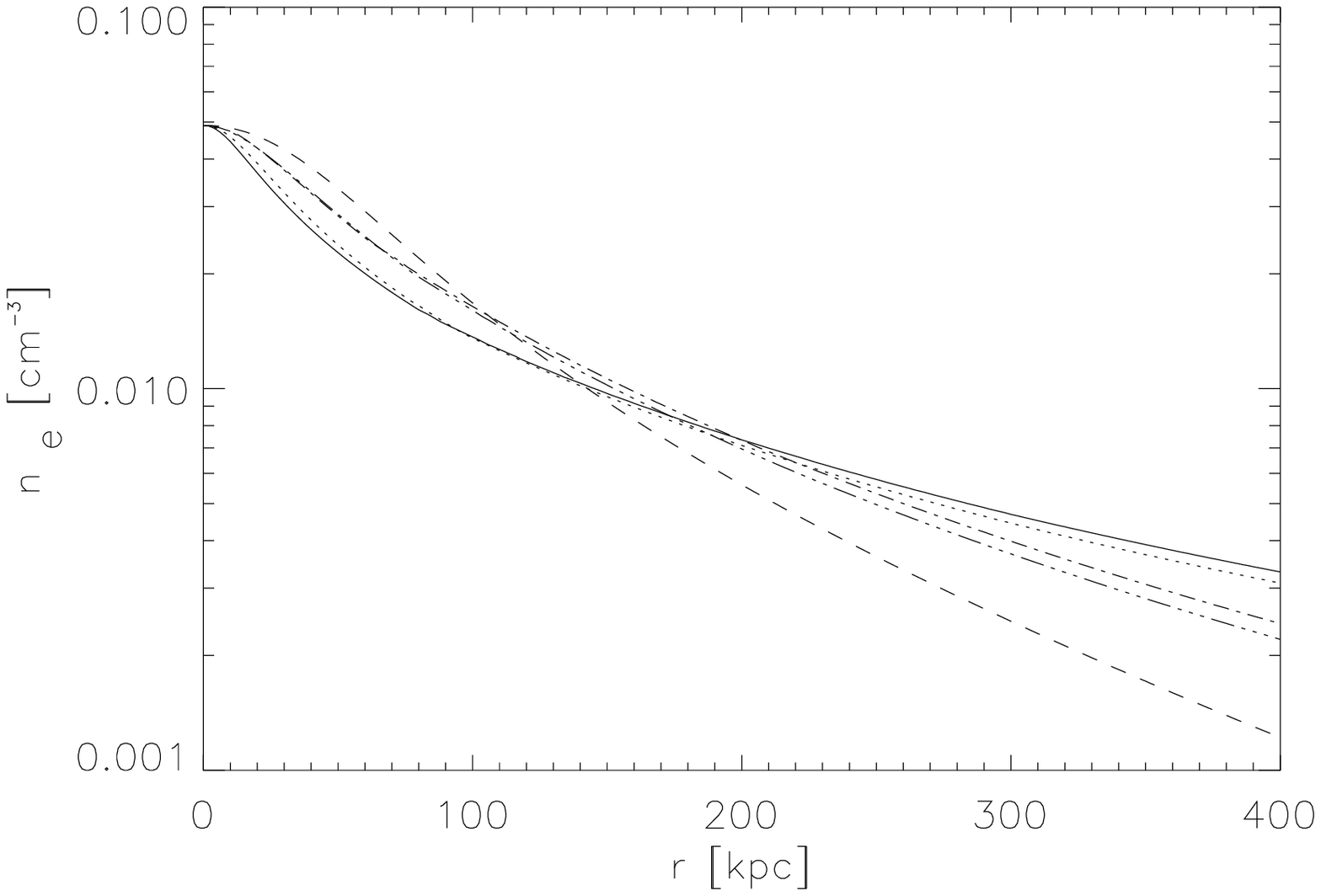} 
\caption{Electron number density profiles corresponding to the models presented in Fig. 3.}
\label{fig4}
\end{figure}

Figures \ref{fig3} and \ref{fig4} are the corresponding figures for a
cluster with a higher central temperature of 4 keV, all other
parameters being the same. For this hotter cluster the luminosity of
the central source has to be higher in order to produce the same
relative effect as in the cooler cluster. In these plots we also
included a model with a high luminosity of $L=3\times 10^{45}$~erg
s$^{-1}$. In this case the temperature distribution is nearly
isothermal.

\begin{figure}[htp]
\plotone{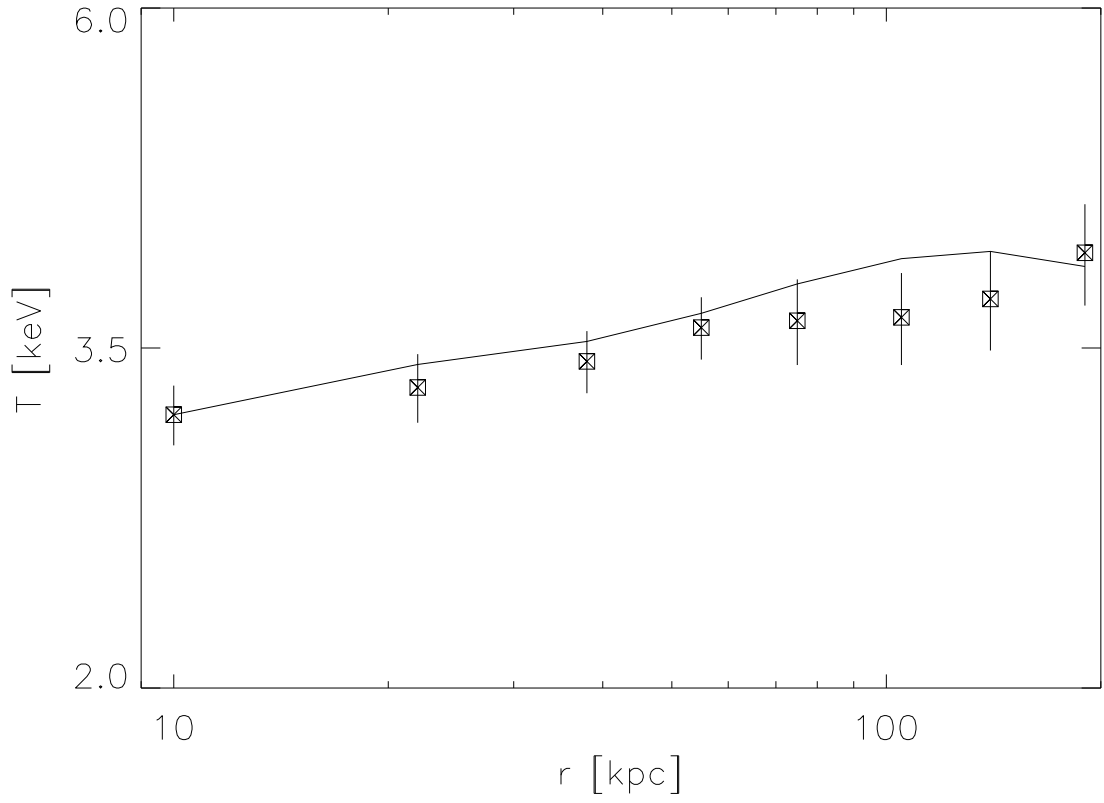} 
\caption{Gas temperature profile of the Hydra A cluster. The crosses
show the results of ACIS data published in \cite{davi01}. The line
corresponds to the fit with $f=0.5$ and $L=4.0\times 10^{45}$ erg
s$^{-1}$.}
\label{fig5}
\end{figure}

Clearly, by including a physically motivated heating term, there are more
degrees of freedom to fit the density and temperature profiles
inferred from observations. This can be demonstrated at the example of
Hydra A which is a well-studied cooling flow cluster with a FR I radio
source (3C218) at its centre. \citet{zak03} have shown that Hydra A
cannot be fitted with a conduction model alone unless one allows for
unplausibly high values for $\kappa$. The NFW parameters for Hydra A
are $r_s=370$ kpc and $M_0=2.9\times 10^{14}\ M_{\odot}$. The central
electron density is $n_e(0)=0.08$ cm$^{-3}$, the temperature at the
centre is $T_{\rm in}=3.11$~keV and at 190 kpc is $T(190\ {\rm
kpc})=4.04$~keV.  In Figure \ref{fig5} we show the temperature profile
as inferred from CHANDRA observations (\citealt{davi01}) together with
our best fit for $f=0.5$ ($L=4.0\times 10^{45}$ erg s$^{-1}$). As is
readily seen, the inclusion of the heating term produces a good fit to
the data without having to invoke an abnormally high conductivity.

A question that we have not addressed so far is the question of the
origin of the energy that is conducted inwards in order to keep the
cluster from collapsing.  If we take a cluster model with no heating,
i.e. $L=0$, and a suppression factor of $f=0.1$ (all other parameters
being those of Fig. \ref{fig1}), the asymptotic value of $r^2\Fc$ at
large radii is $6.2\times 10^{44}$ erg s$^{-1}$. This means that over
a period of 5 Gyrs a total energy of $\sim 10^{62}$ erg has to be
conducted into the cluster which amounts to $\sim 5$ \% of the total
thermal energy of the cluster ($3M_0 k\bar{T}/2\mu m_p$). For higher
values of $f$ this percentage increases. Ultimately, this energy must
be provided by infalling gas at the accretion shock.

It was pointed out by \cite{loeb02} that a significant thermal
conductivity will also lead to energy transport out of the cluster
because beyond the temperature maximum of the cluster the temperature
begins to drop again and the temperature gradient is
reversed. Therefore, conduction is also reversed and heat is
transported from the cluster to its surrounding envelope in these
regions. As estimated by \cite{loeb02}, heat conduction must be
suppressed significantly over the Spitzer value in the outer cluster
regions or else the cores of X-ray clusters would have cooled
significantly over their life times.

\subsection{Stability of the solutions}

We have not performed a stability analysis of the model presented
here. Instead we refer the reader to relevant works that have
been published by others.  The thermal stability of stratified
atmospheres in the presence of radiative cooling and heat conduction
was first analysed by Field (1965) and later generalised by Balbus
(1988) and Balbus and Soker (1989) who performed a Lagrangian
stability analysis. Recently, the issue of thermal instability in
clusters was revisited by \cite{kim03} who examined the global
stability of the models found by \cite{zak03}. On the basis of a
Lagrangian perturbation analysis they concluded that the growth time
of the most unstable global radial mode was $\sim$ 6 - 9 times longer
than the growth time of the local isobaric modes at the centre if the
conductivity was 20 \% to 40 \% of the Spitzer value. Typical growth
times are thus comparable with the time since the last cluster
merger. Hence it is argued that the cluster is thermally stable if
there is a sufficient amount of thermal conduction. The presence of a
heating term as implemented in our model will act to increase the
stability of the cluster.

\subsection{Summary}

We have devised a one-dimensional hydrostatic model for the ICM
where radiative losses are balanced by, both, thermal conduction and
heating by a central source. By including a physically motivated
heating term we have shown that it is possible to fit cluster profiles
without having to invoke unplausibly high values for the thermal
conductivity.

We should concede, however, that our model is very simplified: It
assumes that the cluster is in hydrostatic equilibrium and spherically
symmetric, it does not allow the gas to condense and drop out of the
flow, it ignores magnetic fields and includes a heuristic treatment of
heating by AGN. But despite its simplicity, only a few free parameters
suffice to reproduce the observed profiles of X-ray clusters.\\

\acknowledgments I thank Simon White, Eugene Churazov, Bill Forman and
Christian Kaiser for helpful discussions.

\begin{thebibliography}{}

\bibitem[Afshordi \& Cen(2002)]{afs02} Afshordi, N., \& Cen, R.\ 2002 \apj, 564, 669

\bibitem[Balbus(1988)]{balb88}Balbus, S.A. 1988, \apj, 328, 395

\bibitem[Balbus \& Soker(1989)]{1989ApJ...341..611B} Balbus, S.~A.~\& 
Soker, N.\ 1989, \apj, 341, 611 

\bibitem[Begelman(2001)]{be01} Begelman, M.C. 2001, in {\it Gas and Galaxy
Evolution}, APS Conf. Proc., vol. 240, ed. Hibbard, J.E., Rupen, M.P., and van
Gorkom, J.H., p. 363, (astro-ph/0207656)

\bibitem[Bertschinger \& Meiksin(1986)]{bert86}Bertschinger, E., \&
Meiksin, A. 1986, ApJ, 306, L1

\bibitem[Binney(2001)]{bin01}Binney, J. 2001, 'AGN and cooling flows' in Particles and
Fields in Radio Galaxies, ed. Laing, R. A. , Blundell, K. M., ASP
Conference Series, ASP, San Francisco, in press (astro-ph/0103398)

\bibitem[Bregman \& David(1988)]{breg88}Bregman, J.N., \& David,
L.P. 1988, \apj, 326, 639

\bibitem[Br{\" u}ggen et al.(2002)]{brueg02} Br{\" u}ggen, M., Kaiser,
C.\ R., Churazov, E., Ensslin, E. 2002, \mnras, 331, 545

\bibitem[Br{\" u}ggen \& Kaiser(2002)]{2002Natur.418..301B} Br{\" u}ggen, 
M.~\& Kaiser, C.~R.\ 2002, \nat, 418, 301 

\bibitem[Br{\" u}ggen (2003)]{bru03} Br{\" u}ggen, M. 2003, \apj, submitted

\bibitem[Burns(1990)]{bur90} Burns, J.~O.\ 1990, \aj, 99, 14 

\bibitem[Churazov et al.(2001)], Churazov, E. and Br{\" u}ggen,
M. and Kaiser, C.\ R. and B{\" o}hringer, H. and Forman, W.\ 2001a,
\apj, 554, 261

\bibitem[Churazov et al.(2002)]{chur02}Churazov, E., Sunyaev, R.,
Forman, W., \& B\"ohringer, H. 2002, \mnras, 332, 729

\bibitem[Cowie \& Binney(1977)]{1977ApJ...215..723C} Cowie, L.~L.,
Binney, J.\ 1977, \apj, 215, 723

\bibitem[David et al.(2001)]{davi01}David, L.P., Nulsen, P.E.J.,
McNamara, B.R., Forman, W., Jones, C., Ponman,T., Robertson, B., \&
Wise, M. 2001, \apj, 557, 546

\bibitem[Fabian \& Nulsen(1977)]{1977MNRAS.180..479F} Fabian, A.~C.,
Nulsen, P.~E.~J.\ 1977, \mnras, 180, 479

\bibitem[Fabian(1994)]{1994ARA&A..32..277F} Fabian, A.~C.\ 1994,
\araa, 32, 277

\bibitem[Fabian et al.(2001)]{fabi01}Fabian, A.C., Mushotzky, R.F.,
Nulsen, P.E.J., \& Peterson, J.R. 2001, \mnras, 321, L20

\bibitem[Field(1965)]{fiel65}Field, G.B. 1965, \apj, 142, 531

\bibitem[Friaca(1986)]{fri86} Friaca, A.~C.~S.\ 1986, \aap, 
164, 6 

\bibitem[Gruzinov (2002)]{gruz02}Gruzinov, A. 2002, astro-ph/0203031

\bibitem[Kim \& Narayan(2003)]{kim03} Kim, W.-T., \& Narayan, R. 2003,
\apj, submitted (astro-ph/030397)

\bibitem[Loeb(2002)]{loeb02}Loeb, A. 2002, New Astr., 7, 279

\bibitem[Malyshkin \& Kulsrud(2001)]{maly01}Malyshkin, L., \& Kulsrud,
R. 2001, \apj, 549, 402

\bibitem[Maoz et al.(1997)]{mao97}
   Maoz, D., Rix, H.-W., Gal-Yam, A., \& Gould, A.\ 1997, \apj, 486, 75

\bibitem[Markevitch et al.(2000)]{mark00}Markevitch, M. et al. 2000,
\apj, 541, 542

\bibitem[Markevitch(2002)]{mark02}Markevitch, M. 2002 (astro-ph/0205333)

\bibitem[Medvedev \& Narayan(2002)]{medv02} Medvedev, M.V., \& 
Narayan, R. 2002, \mnras, submitted (astro-ph/0107066)

\bibitem[Meiksin(1988)]{meik88} Meiksin, A.\ 1988, \apj, 334, 
59

\bibitem[Narayan \& Medvedev(2001)]{nara01}Narayan, R., \& Medvedev,
M.V. 2001, \apj, 562, L129

\bibitem[Navarro et al.(1997)]{nava97}Navarro, J.F., Frenk, C.S., \&
White, S.D.M. 1997, \apj, 490, 493

\bibitem[Owen et al.(2000)]{oek00}
Owen F.N., Eilek J.A., Kassim N.E., 2000 \apj, submitted, astro-ph/0006150

\bibitem[Quilis, Bower, \& Balogh(2001)]{qui01} Quilis, V., 
Bower, R.~G., \& Balogh, M.~L.\ 2001, \mnras, 328, 1091 

\bibitem[Rechester \& Rosenbluth(1978)]{rech78}Rechester, A.B., \&
Rosenbluth, M.N. 1978, Phys. Rev. Lett., 40, 38


\bibitem[Rybicki \& Lightman(1979)]{rybi79}Rybicki, G.B., \&
Lightman, A.P. 1979, Radiative processes in astrophysics (New York:
Wiley)

\bibitem[Ruszkowski \& Begelman(2002)]{rus02} 
   Ruszkowski, M., \& Begelman, M.\ C.\ 2002, \apj, 581, 223

\bibitem[Sarazin(1988)]{1988cfcg.work....1S} Sarazin, C.~L.\ 1988,
NATO ASIC Proc.~229: Cooling Flows in Clusters and Galaxies, Kluwer,
Dordrecht, 1

\bibitem[Spitzer(1962)]{spi62} 
   Spitzer, L.\ 1962, Physics of Fully Ionized Gases (New York: Interscience)

\bibitem[Sutherland and Dopita(1993)]{sd93} Sutherland, R.S., \& Dopita, 
M.A. 1993, ApJS, 88, 253

\bibitem[Tabor \& Binney(1993)]{tabo93}Tabor, G., \& Binney, J. 1993,
\mnras, 263, 323

\bibitem[Vikhlinin et al.(2001a)]{vikh01a}Vikhlinin, A., Markevitch,
M., \& Murray, S. S. 2001a, \apj, 549, L47

\bibitem[Vikhlinin et al.(2001b)]{vikh01b}Vikhlinin, A., Markevitch,
M., \& Murray, S. S. 2001b, \apj, 551, 160

\bibitem[Voigt et al.(2002)]{voig02}Voigt, L.M., Schmidt, R.W.,
Fabian, A.C., Allen, S.W., \& Johnstone R.M., 2002, \mnras, submitted
(astro-ph/0203312)

\bibitem[Voit \& Bryan(2001)]{voit01} Voit, G.~M.~\& Bryan, 
G.~L.\ 2001, \nat, 414, 425

\bibitem[Zakamska \& Narayan(2003)]{zak03}
   Zakamska, N.\ L., \& Narayan, R.\ 2003, \apj, 582, 162

\end{thebibliography}
\end{document}